# Floquet Topological Insulators for Sound

Romain Fleury[1,*], Alexander B. Khanikaev[2,3,*], Andrea Alù[1]

[1] Department of Electrical and Computer Engineering, The University of Texas at Austin, 1616 Guadalupe St., Austin, TX 78701, USA

[2] Department of Physics, Queens College of The City University of New York, Queens, NY 11367, USA

[3] Department of Physics, The Graduate Center of The City University of New York, New York, NY 10016, USA

*The unique conduction properties of condensed matter systems with topological order have recently inspired a quest for similar effects in classical wave phenomena. Acoustic topological insulators, in particular, hold the promise to revolutionize our ability to control sound, allowing for large isolation in the bulk and broadband one-way transport on their edges, with topological immunity against structural defects and disorder. So far these fascinating properties have been obtained relying on moving media, which inherently introduce noise and absorption losses, hindering the practical applicability of topological acoustics. Here, we overcome these limitations by modulating in time the acoustic properties of a lattice of resonators, introducing the concept of acoustic Floquet topological insulators. Acoustic modulation at ultrasonic frequencies can reach values as high as a few percents, enabling broadband effects with an on-site, moderate modulation strategy that surprisingly does not require phase uniformity across the lattice. Using first-principle numerical experiments, we show that acoustic waves provide a fertile ground to apply the anomalous physics of Floquet topological insulators, and demonstrate their relevance for a wide range of acoustic applications, including broadband acoustic isolation and topologically protected, nonreciprocal acoustic emitters.*

[*]These authors equally contributed to the paper

In most acoustic media, time-reversal symmetry requires that sound travels on two-way channels: if a wave can propagate in a given direction, propagation in the opposite direction is also allowed. Therefore, acoustic modes come in pair, and any forward propagating state is always associated with a backward state at the same frequency. Defects or interfaces that couple forward and backward waves are thus inherently associated with reflections, leading to the central issue of impedance matching, a pivotal problem in the engineering and design of acoustical systems.

The seemingly unrelated field of condensed matter physics offers new solutions to tackle the challenges associated with acoustic impedance matching. Recent years have witnessed the discovery of fermionic condensed matter systems characterized by a unique type of order, topological in nature [1-10]. This discovery did not only expand the existing classification of solid-state systems, but also significantly enriched our understanding of quantum and classical phenomena in different branches of physics [11-15]. Topological condensed-matter systems feature robust unidirectional bandgap-crossing edge states, offering unusual conduction properties. These edge states exhibit immunity to a broad range of structural imperfections, inherently avoiding backscattering over broad energy ranges, and circumventing localization in the presence of disorder [16]. Interestingly, while originally proposed in the context of quantum fermionic condensed matter systems, the concept of topological order also opens a wealth of new possibilities when extended to classical waves [14,15,17-21] and bosons [13,22-26]. In these systems, which lack the protection available for fermions, topological order emerges from the removal of either time-reversal symmetry [13,17,18,20,27-30], emulating the Quantum Hall effect, or of certain forms of internal or lattice symmetries [19,21,31-36], which is in analogy with topological crystalline insulators in condensed matter [37]. However, only the first method, breaking time-reversal symmetry, can guarantee the absence of reflected modes regardless of the nature of the defect. In acoustics, indeed, topological insulators with strong topological protection against defects and disorder have been obtained in time-asymmetric sonic crystals, such as networks of acoustic cavities filled with a fluid in motion [14,15]. This solution, however, may arguably be unpractical, since implementing uniform motion in a lattice is challenging, and the inherent losses and noise that intrinsically accompany acoustic propagation in moving media are largely detrimental in most application scenarios.

In this article, we solve these issues and enrich the toolkit of acoustic physicists and engineers introducing a nonreciprocal material platform possessing the unique ability to eliminate reflections and impedance matching challenges for sound. This is obtained in an acoustic lattice whose properties are modulated in space and time in a time-harmonic and rotating fashion, demonstrating an acoustic analogue of Floquet topological insulators [38-44]. Different from time-modulated lattices proposed in the context of photonics [29,30,44], our proposal is based on a slow, on-site rotating modulation scheme, which remarkably, as we prove in the following, does not require at all phase uniformity between different lattice sites. This allows us to take a pivotal step towards practical applications of topological insulators in acoustic systems. We exploit the fact that acoustic properties can be modulated in a strong fashion, up to tens of percents, compared to electromagnetic properties, opening the possibility to broadband, topologically protected, one-way acoustic devices, including an ultrabroadband acoustic diode that transmits sound only in one direction, and a topologically protected acoustic emitter.

While the concept of Floquet topological insulators was originally introduced in solids [38-43], in photonics [20,29,30,44], and recently generalized to the case of photo-elastic systems [13], the theoretical methods used to demonstrate this class of topological order have so far been limited to abstract tight binding and other idealized models. Thus, even on the theoretical front, Floquet topological order in classical systems has until now evaded rigorous treatment based on first-principle theory and numerical simulations. Here on the contrary we develop and apply a rigorous full-wave treatment, based on first principles, to demonstrate the realistic possibility of acoustic Floquet topological insulators and its application to practical devices and new concepts for sound engineering.

The system we propose and investigate in the following is shown in Figure 1a. Consider an acoustic crystal with hexagonal lattice formed by coupled acoustic trimers, connected together along the hexagonal bonds via small rectangular channels. Each trimer can be viewed as a resonant acoustic metamolecule composed of three acoustic cavities coupled by cylindrical waveguides [45]. The acoustic medium filling the crystal is silicon rubber RTV-602, an ultra-low loss material widely used in ultrasonic systems, with density $\rho_0 = 990 \, \text{kg/m}^3$ and compressibility $\beta_0 = 9.824 \; 10^{-10} \, \text{Pa}^{-1}$ [46]. The surrounding medium is air, and the cavity diameter is 1 cm with a thickness of 3mm, supporting the first dipolar cavity resonance around 60 kHz. In the following, we operate well below this cavity resonance, in a frequency range in which each cavity can be modeled as a lumped element storing a net amount of acoustic pressure, and described by its acoustic capacitance $C_0 = \beta_0 V_0$ [47]. A trimer formed by three cavities is therefore equivalent to a $L-C$ resonating loop, supporting a doubly-degenerate lumped resonance at 21.6 kHz associated with the resonant exchange of potential and kinetic energy among the cavities and the internal coupling channels. The trimers are weakly coupled to each other and used as building blocks of the hexagonal lattice. In order to break time-reversal symmetry in this resonant acoustic metamaterial and induce topologically non-trivial properties, we modulate the acoustic capacitance of each cavity by a time-dependent on-site potential $\Delta C_m(t) = \delta C \cos(\omega_m t - \varphi_m)$, enforcing a time-harmonic modulation with strength $\delta C / C_0$ and frequency $f_m = \omega_m / 2\pi$. The phase $\varphi_m$ depends on the considered sub-cavity *m*, so that the modulation imparts an effective spin onto each trimer, breaking time-reversal symmetry. This modulation protocol is first assumed to be uniform within the crystal, as summarized in Figure 1b, which sketches the modulation strategy over each unit cell. This may be easily achieved in practice by compressing the volume of each cavity using piezoelectric actuators. Modulation up to several tens of percents are possible using suitable actuating strategies, which may be leveraged to further increase the bandwidth of the topologically nontrivial bandgap discussed in the following.

The effect of a weak spatio-temporal modulation on the bulk band structure of the acoustic crystal is shown in Figure 1c. In the absence of modulation (blue curves), four propagating bands are found around 22 kHz, along with two high frequency bands near 56.5 kHz (not seen in the figure), consistent with the fact that the lattice is formed by two metamolecules per unit cell, each providing three degrees of freedom associated with the lowest order mode of the isolated cavities. Here we focus on the four low frequency bands possessing a dipolar pressure-field profile (corresponding to angular momentum $l = +1, -1$), which perceive the effective angular-momentum bias [48,49] provided by the chosen form of spatio-temporal modulation. Two of these bands appear to be slow in nature, corresponding to "deaf" modes with vanishingly small group velocity, whereas the other two correspond to fast Dirac bands, with degeneracy at both Γ and K points, due to the time-reversal T and combined parity–time-reversal PT symmetry properties of the lattice, respectively. When modulation is applied (orange curves, obtained for $\delta C / C_0 = 6\%$, $f_m = 2\,\text{kHz}$), the band structure folds along the frequency axis, with periodicity equal to $\omega_m$, consistent with Floquet-Bloch theorem in the time domain. Given the breaking of temporal symmetry, degeneracy is lifted both at Γ and K points by an amount proportional to the modulation depth.

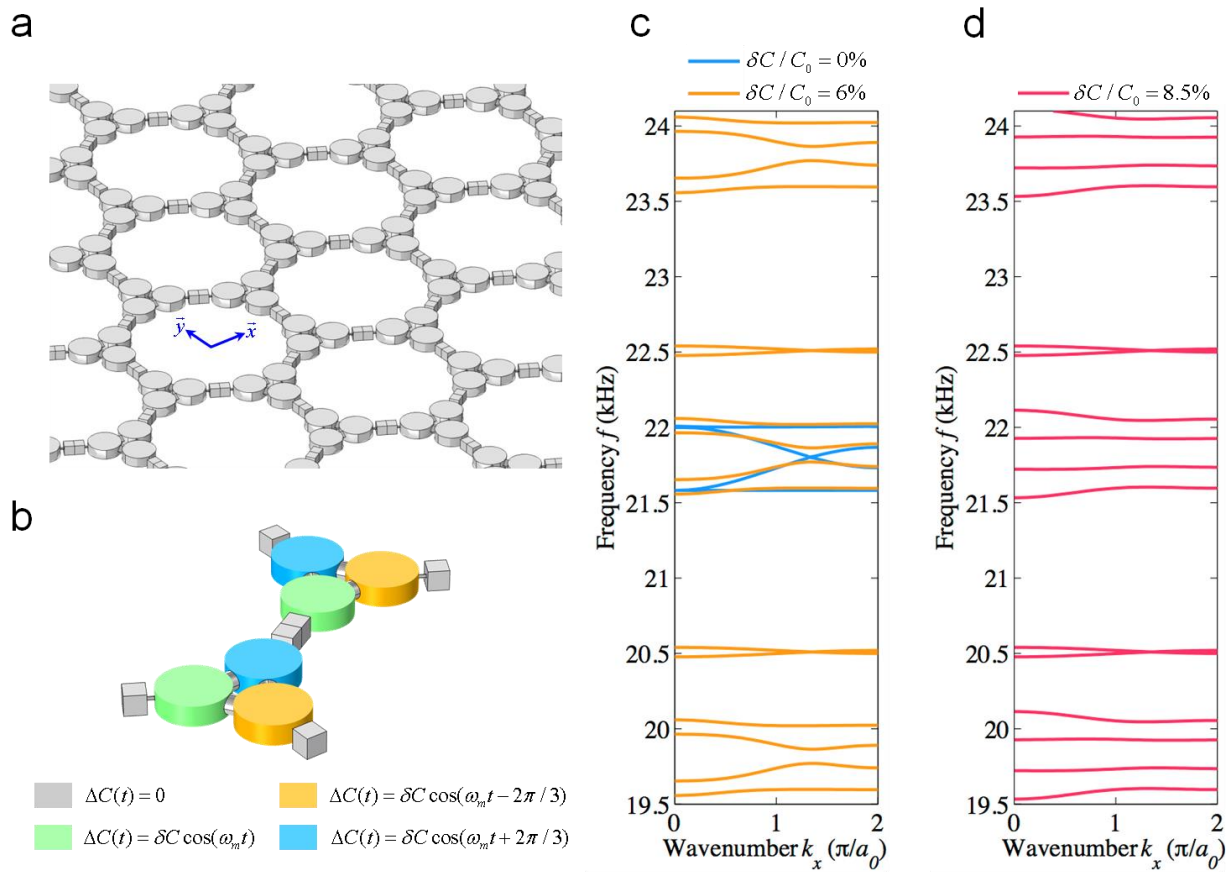

Figure 1: **Hexagonal lattice of modulated acoustic trimers forming an acoustic Floquet topological insulator.** (a) We consider a time-dependent phononic crystal based on a hexagonal lattice of acoustic trimers. The trimers are composed of three interconnected cylindrical cavities (1 cm in diameter), coupled to each other via small rectangular channels. The grey material in the figure is silicon rubber, a water-like ultrasonic material with large density contrast with surrounding air. In each unit cell (b), the acoustic capacitance $C = C_0 + \Delta C(t)$ of each cavity is periodically modulated in time at a frequency $\omega_m$ and with amplitude $\delta C$, in a rotating fashion, with uniform handedness throughout the lattice. (c) Comparison between the bulk band structure in absence of modulation (blue curves) and in the presence of modulation (orange curves, $\delta C / C_0 = 6\%$, $f_m = 2\,\text{kHz}$). The time modulation has the effect of folding the band structure along the frequency axis and lifting the degeneracy at the K and **Γ** points, opening a complete bandgap with topological protection. (d) Optimal gap opening occurs at $\delta C / C_0 = 8.5\%$ for $f_m = 2\,\text{kHz}$.

Notably, this happens only for dipolar bands, and the high frequency bands possessing double degeneracy at K-point remain unaffected by the modulation, due to their monopolar ($l = 0$) nature. When the modulation depth is further increased, with $f_m$ unchanged, the bandgap between the two inner bands further opens until the two bands can become flat. This special condition, shown in Figure 1d (red curves), occurs for $\delta C / C_0 = 8.5\%$, and it corresponds to the maximal gap bandwidth. Further increase in modulation depth is detrimental to the gap opening, due to pinching of the bands at the $\Gamma$ point. It is important to note that we operate here around the lattice resonance, for which the coupling constant between trimers $J$, whose order of magnitude is given by the spectral span of the unmodulated (blue) band structure (a few hundreds Hz), is smaller than the modulation frequency $f_m$ (2 kHz). This ensures that no resonant transitions between different Floquet bands can take place, and bands of different origin and different Floquet orders remain isolated from each other. At the same time, this condition ensures that the effect of on-site modulation is enhanced by the resonant nature of the trimers, enabling strong breaking of time-reversal symmetry with a modulation of a few percents only, at a frequency (2 kHz) one order of magnitude smaller than the acoustic frequency (20 kHz).

The appearance of a band gap in the spectrum of the driven system, demonstrated here by first-principle simulations, does not guarantee its topological nature. A more detailed analysis of the bands and corresponding eigenmodes of the system is required to determine the class of topological order. In general, temporal modulation of two-dimensional systems can lead to topological order of two distinct classes, both supporting edge states: one characterized by a non-vanishing topological invariant of the first Chern class, known as a Chern insulator (CI), the other one with a vanishing Chern number, or anomalous Floquet insulators (AFI), when different Floquet orders interact leading to an exchange of topological charges [41,43]. Since our temporal modulation scheme clearly violates time-reversal symmetry, and since the modulation is sufficiently weak to not resonantly couple different Floquet orders, we can expect the opened bandgap to have topological nature of the first Chern class [41]. In order to fully reveal the topological phase induced by the temporally modulated system of Fig. 1, we developed a semi-analytical approach based on the derivation of the effective Hamiltonian of the time-dependent tight-binding model [41], which allows a traditional characterization of the system in terms of topological invariants. The geometry under analysis can be exactly mapped onto a tight-binding Hamiltonian associated to a hexagonal lattice with nearest neighbor hopping, which should be modified to account for the internal structure of the metamolecules

$$\hat{H}(t) = \sum_{m} \hat{\varepsilon}_m(t)|m\rangle\langle m| + \sum_{m,n}(\hat{J}|n\rangle\langle m| + \hat{J}^*|m\rangle\langle n|), \tag{1}$$

where $|\boldsymbol{m}\rangle = |m_1, m_2, m_3\rangle$ is a vector whose three components correspond to the three acoustic cavities forming each node of the hexagonal array. The time-independent hopping terms are given by a diagonal 3x3 matrix $\hat{J}_m = J\hat{I}$, where $J$ measures the coupling strength between neighboring cavities of adjacent metamolecules, while the time-dependent on-site term assumes the form

$$\hat{\varepsilon}_m = \begin{pmatrix} \omega_0 + \delta\omega_1(t) & \kappa & \kappa^* \\ \kappa^* & \omega_0 + \delta\omega_2(t) & \kappa \\ \kappa & \kappa^* & \omega_0 + \delta\omega_3(t) \end{pmatrix}, \tag{2}$$

where $\delta\omega_l(t) = \delta\omega \cos(\omega_l t + \phi_l)$, and $\phi_l = \frac{2\pi}{3}l$. The meaning of the parameters introduced here directly follows from the form of the Hamiltonian: $\omega_0$ is the on-site energy, i.e., the lowest resonance frequency of the isolated acoustic cavities, $\delta\omega_l(t)$ is its temporal modulation, and $\kappa$ is the on-site coupling between cavities of the same metamolecule. We discuss this model in more details in the Methods. The Hamiltonian (1) defines the evolution operator $\widehat{U}(t) =$

$\mathcal{T}\exp\left[-i\int_0^t \widehat{\mathcal{H}}(t)\,dt\right]$, where $\mathcal{T}$ is the time-ordering operator. After writing the Hamiltonian (1) in reciprocal space, we can introduce introduce a stroboscopic evolution operator $\widehat{U}(T)$ on every period of the system $t = nT$, and determine the topological class of the system from its time-independent "effective" Hamiltonian [41] $\widehat{\mathcal{H}}_{eff} = \frac{i}{T}\log \widehat{U}(T)$ . We extracted the effective parameters in Eqs. (1-2) directly from first-principle FEM simulations to calculate the Chern numbers $C_n = \frac{1}{2\pi}\int_{BZ}(\partial_{k_x}A_y - \partial_{k_y}A_x)\,d^2\mathbf{k}$ for every band $n$ using the numerically calculated eigenstates $|p_n\rangle$ of the effective Hamiltonian $\widehat{\mathcal{H}}_{eff}$ and the Berry connection $\mathbf{A} = -i\langle p_n|\partial_{\mathbf{k}}|p_n\rangle$. After integration over the entire Brillouin zone, we found that the four bands of interest possess topological indices $C_n = \{1, 0, 0, -1\}$ and, as expected, reversal of the modulation scheme from clockwise to counterclockwise, equivalent to time-reversal, causes the reversal of the Chern numbers $C_n = \{-1, 0, 0, 1\}$ (see Methods).

One of the most appealing features of topological Chern insulators, including the one proposed here, is the existence of one-way edge modes at the boundaries between domains of different topology. The number of topological edge states supported by a given interface is dictated, according to the bulk-boundary correspondence principle [50], by the change of sum of Chern numbers of all the bulk bands of lower frequency across the interface [51]. In Fig. 2 we consider edge modes propagating on the external edges of a crystal with uniform modulation handedness, as well as edge modes that propagate along the boundary between two crystal domains with opposite modulation handedness. In the first case, the considered interface involves a topologically trivial half space, and therefore we expect only one edge mode, regardless of the crystal termination. The domain wall, for which the difference in the sum of Chern numbers equals two, is expected to support two topologically nontrivial acoustic edge modes.

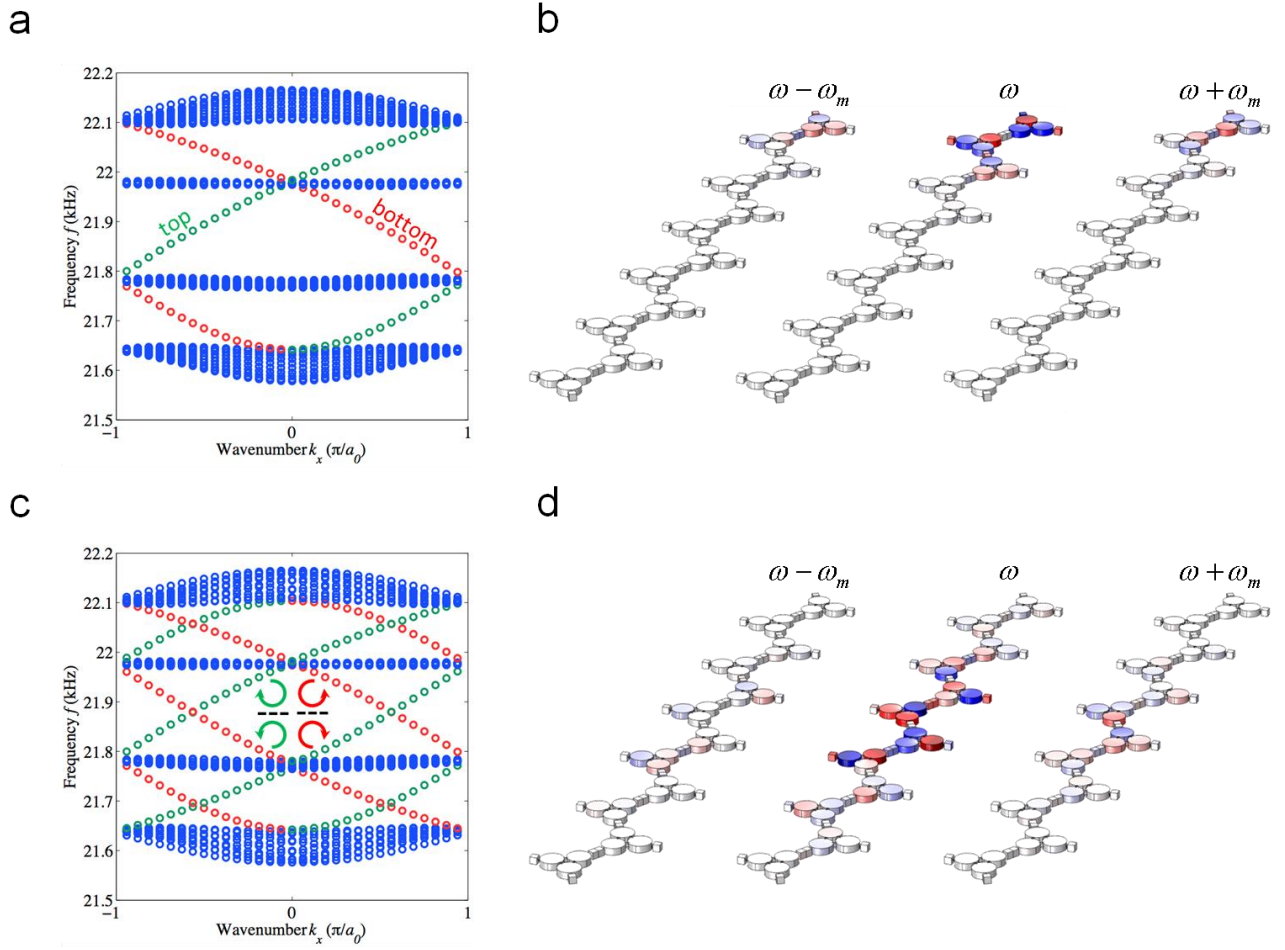

Figure 2: **Topological edge modes of acoustic Floquet topological insulators.** (a) Acoustic band structure for a supercell composed of a 1 by 6 array of unit cells, terminated by a hard wall boundary at the top and bottom, and periodic boundary conditions along $x$. Blue dots refer to bulk modes. Green dots refer to edge modes localized at the top boundary of the supercell. Red dots refer to edge modes localized at the bottom boundary of the supercell. (b) Acoustic pressure field distribution for an edge mode localized at the top boundary, showing three harmonic components: $\omega - \omega_m$ (left), $\omega$ (center) and $\omega + \omega_m$ (right). (c) Acoustic band structure for a supercell composed of a 1 by 6 array of unit cells supporting a domain wall at the center: the modulation of the three unit cells at the bottom is right-handed, and at the top it is left-handed. The supercell is surrounded by periodic boundary conditions in all directions. Blue dots refer to bulk modes. Red and green dots refer to edge modes localized at boundaries with opposite handedness, as represented in the inset. (d) Same as (b), but in the case of a domain wall boundary at the supercell center.

Figure 2a considers the case of a crystal terminated by a hard-wall boundary condition, which naturally occurs if the crystal is surrounded by an acoustic medium with very different density, such as air. The figure represents the band structure obtained for a 1x6 supercell, modulated with $\delta C/C_0 = 8.5\%$ and $f_m = 2\,\text{kHz}$. The supercell is terminated at the top and bottom by hard wall boundary conditions, with periodic boundary conditions along *x*. Figure 2a focuses on the frequency region of the four bulk bands of interest: the blue dots correspond to bulk modes, which form four bands separated by a gap, the two in the middle being flat, consistent with Figure 1c. Differently from Figure 1, however, the truncated structure now supports two distinct modes within each bandgap. Inspection of the mode profile shows that these modes, with positive (green dots) and negative (red dots) group velocity, are localized at the top and bottom boundaries of the supercell, respectively, thus confirming the one-way response and the chiral character of the modes, which is dictated by the modulation handedness of each unit cell. Interestingly, the edge modes are characterized by different Floquet harmonics, due to the effect of modulation, all confined along the edge. Figure 2b shows the acoustic pressure distribution of the three dominant frequency harmonics ($\omega - \omega_m$ (left), $\omega$ (center), $\omega + \omega_m$ (right)) for the edge mode localized at the top boundary.

Figure 2b considers the case of a domain wall – the interface between two crystal domains with opposite modulation spin. The plot represents the band structure of a 1x6 supercell with the bottom three cells and top three cells having opposite modulation handedness. The supercell is surrounded by periodic boundary conditions in all directions. According to the bulk-boundary correspondence principle [50], this sudden change of modulation spin should be associated with a pair of acoustic edge modes in each band gap region. In addition, the "time-reversed" boundary effectively emerging because of the periodic boundary condition at the top and bottom of the supercell supports another pair of acoustic edge modes. A total of four acoustic edge modes are therefore expected in each bandgap region for this case, as confirmed by the band structure in

Figure 2b. Four chiral edges modes that appear within each topological bandgaps can be classified according to the sign of their group velocity: the green dots correspond to positive group velocities, while the red dots correspond to negative group velocities. Inspection of the modal profiles confirms that the modes with positive (respectively negative) group velocity along $x$ are associated with a right handed (RH) to left handed (LH) modulation flip boundary (respectively LH to RH flip), as the boundary is crossed in the $y$ direction. This confirms the prediction that a single boundary hosts two edge states with unidirectional properties. Figure 2c shows the pressure field distribution associated with the edge mode localized at the center boundary for all three dominant harmonics.

Breaking time-reversal symmetry in classical and bosonic topological insulators allows for strong topological protection against defects and local disorder [18]. The absence of backward propagating modes indeed ensured absence of back-reflections due to defects and reflections at the load. To confirm the topological robustness of the Floquet acoustic edge modes, we performed large-scale numerical experiments, deliberately introducing different kinds of lattice defects. Figure 3a shows a drastic example of an acoustic edge state that seamlessly flows along a crystal edge despite the presence of sharp cuts and stringent turns in the hexagonal lattice, and abrupt transitions from zigzag, to armchair, to bearded edge types. Even a severe lattice defect in the form of a missing trimer, placed at the worst location for the edge mode (i.e., right on the edge), does not affect one-way propagation. In any natural acoustic material, with topologically trivial properties, the defects would inherently couple forward and backward waves leading to reflections. On the contrary, the chiral acoustic edge states are backscattering immune and exhibit strong protection against any kind of defect, providing robust and reconfigurable propagation over a broad bandwidth. Figure 3b shows the case of a chiral mode localized inside the cavity which is dual to that localized at an external edge. Unlike modes localized to a defect in a topologically trivial bandgap material, such a topologically protected spinning mode can only carry positive angular momentum, and does not depend on the cavity shape or termination type, but solely on the cavity perimeter and the topological properties of the surrounding lattice.

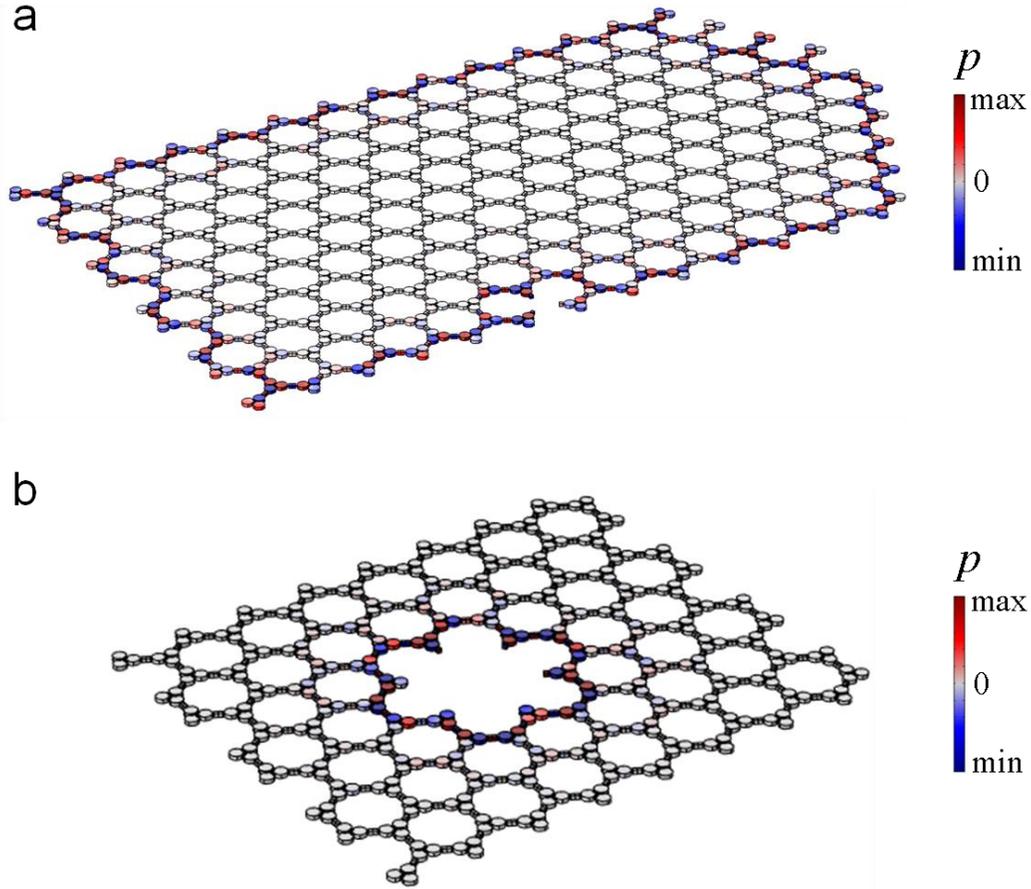

Figure 3: **Strong topological protection of acoustic chiral modes against defects.** (a) Immunity of the non-reciprocal edge mode to various defects and turns. The one-way edge states flows seamlessly regardless of the edge geometry, shape of the turn, or the presence of localized defects. (b) Example of uni-rotational acoustic mode localized to a crystal defect in an inner edge.

Remarkably, the acoustic edge states supported by acoustic Floquet topological insulators are not only robust against structural defects, but also to disorder in the modulation scheme, whose phase can be arbitrary from site to site. We demonstrate this extraordinary property using the input-output numerical experiment shown in Figure 4. First, we consider a perfectly ordered modulation scheme, represented in Figure 4a (top). The subcavities composing a given trimer are modulated with $2\pi/3$ phase shifts, and the modulation scheme is homogeneous from trimer to trimer. This ideal situation is compared to the largely disordered scenario of Figure 4b (top) in which each trimer is still modulated in a rotating fashion, but an arbitrary random phase is added from trimer to trimer. The bottom plots show that the acoustic pressure distribution in each case is identical both in magnitude and in phase, demonstrating that topological protection is not affected at all by large modulation disorder within the lattice. It is only the handedness of the modulation within trimers that defines the topological properties of the crystal. Thus, our on-site modulation scheme does not have at all to be uniform throughout the crystal, contrary to previous works on time-dependent photonic lattices [29,30,44] that are based on off-site modulation of coupling terms, and require stringent constraints of modulation phase uniformity across large distances. With this remarkable property, our proposal constitutes a key step towards the practical implementation of robust, large-scale, acoustic Floquet topological insulators.

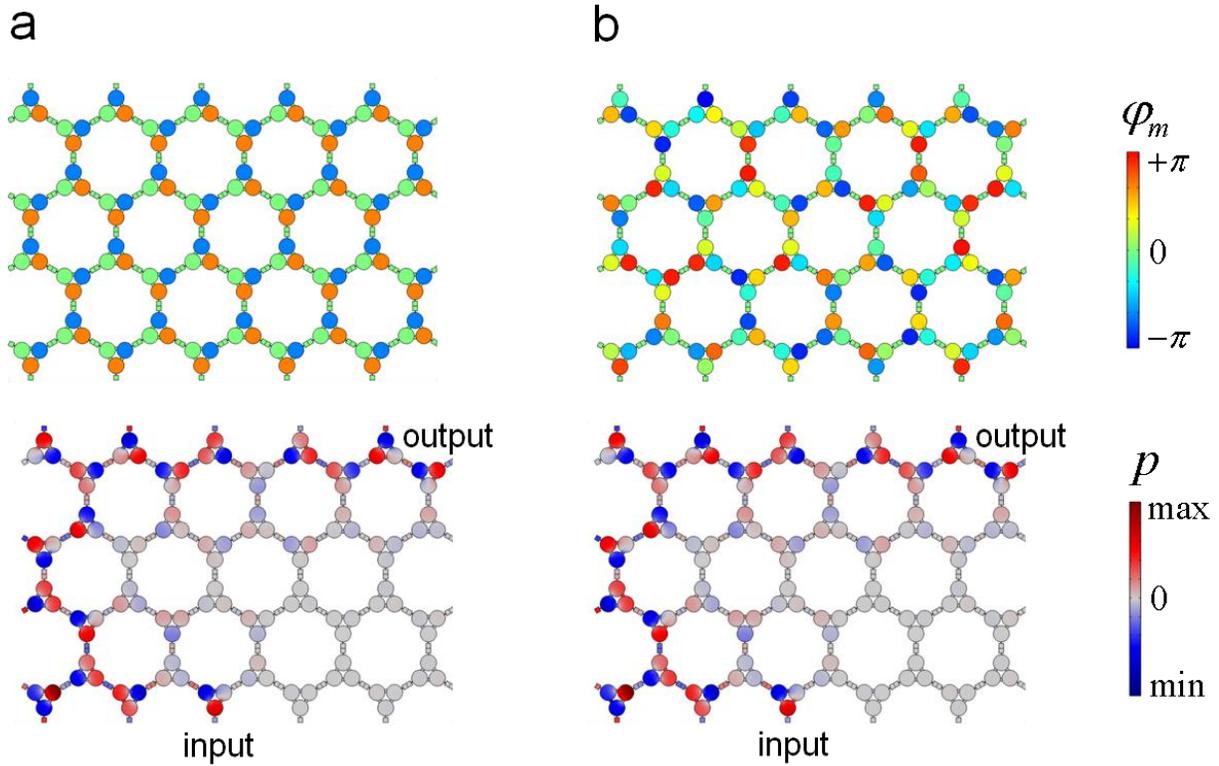

Figure 4: **Robustness to modulation phase disorder.** (a) When the modulation is uniform within the crystal (top), each trimer is driven with the same phase, and transmission between two points via the edge mode is topologically protected (bottom). (b) Remarkably, this behavior is not affected if each trimer is independently modulated with a random phase, and the same acoustic edge mode is excited, both in amplitude and phase.

In addition to robustness against defects and irregularities in geometry or modulation scheme, the proposed acoustic Floquet topological insulators exhibit unique waveguiding properties that enable unprecedented control over wave propagation, leading to a plethora of new potential applications. As seen in Figure 5a, by tailoring line boundaries of the domain walls within a lattice with opposite modulation spin on each side of the boundary, it is possible to create dynamically reconfigurable, backscattering-immune, broadband non-reciprocal waveguides. Here, we have imparted a domain wall boundary to control reflectionless routing of an acoustic edge mode between port 1 and port 2. Inside the crystal, the mode excited at port 1 travels seamlessly to port 2 along the domain wall, regardless of sharp turns and defects along its irregularly shaped path, confirming the possibility to dynamically control and route acoustic signals by controlling the modulation spin, avoiding back-reflections. Remarkably, this non-reciprocal reconfigurable waveguide can be used to build a broadband acoustic diode, i.e., a device which transmits acoustic waves in the forward direction $1 \to 2$, but not in the reverse direction $2 \to 1$, over a large range of frequencies. Indeed, as seen in Figure 5b, the forward acoustic power transmission coefficient $T_{1 \to 2}$ (blue solid curve) is unitary over a wide frequency range, only vanishing at a few isolated frequency points corresponding to the bulk crystal bands (blue points in Fig. 2b), whereas transmission in the reverse direction $T_{2 \to 1}$ (red solid curve) is always negligible. This wideband diode-like behavior is not a trivial effect that one can attribute to the non-reciprocal response of a single trimer, which would be inherently limited to a single frequency point [45,48], but rather to a complex lattice effect intimately related to the topologically non-trivial nature of wave propagation. To illustrate this point, we plot the transmission coefficients $T_{1 \to 2}$ and $T_{2 \to 1}$ for a single trimer for comparison (shown in the inset). It is clear that, even if the single trimer is capable of largely breaking reciprocity, it only allows to do so over a bandwidth of zero measure, i.e., at a single frequency, and with $T_{1 \to 2} < 1$. Conversely, when the trimers are assembled in a lattice with a domain wall type of topological boundary, collective effects induce perfect ($T_{1 \to 2} = 1$, $T_{2 \to 1} = 0$) diode-like behavior over a wide, and customizable frequency range. Unlike previous proposals on non-reciprocal acoustics [52,53], the non-trivial topological properties of the lattice induce ideal acoustic diode functionality over a large bandwidth.

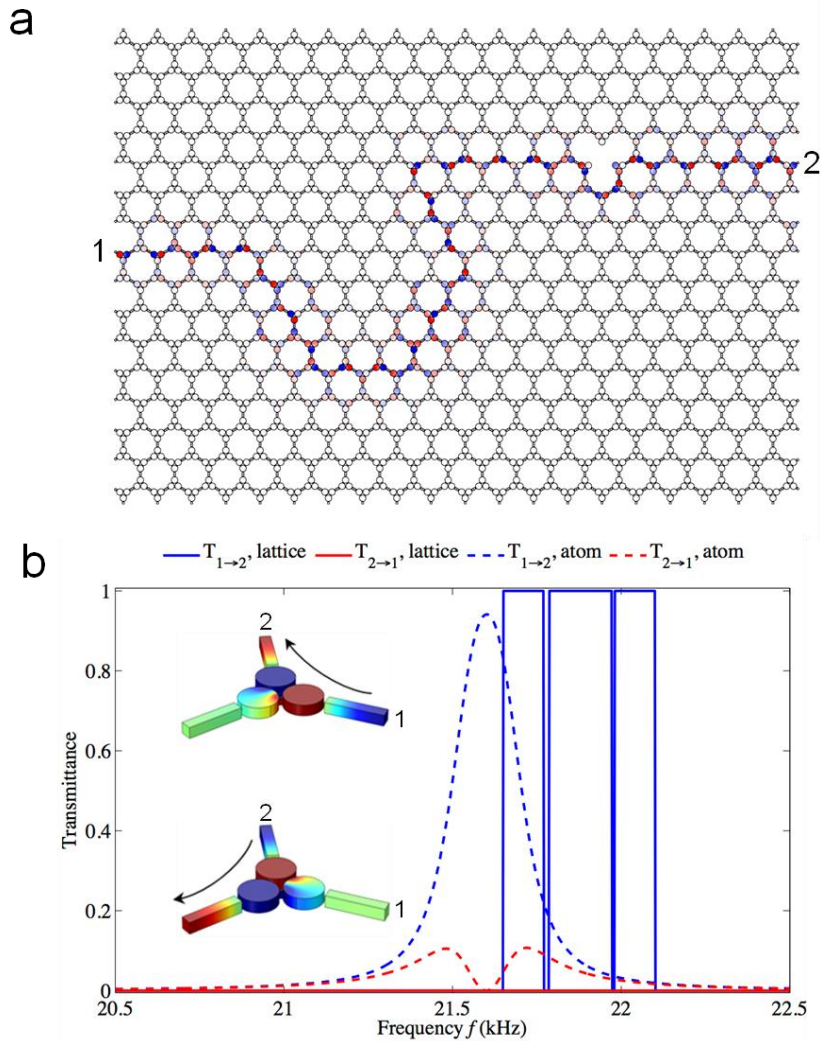

Figure 5: **Broadband topologically protected acoustic diode.** (a) The boundary between two crystal domains with opposite modulation handedness makes a reconfigurable, backscattering-immune, and broadband non-reciprocal waveguide, that perfectly transmits acoustic waves along the boundary, from the input port 1 to the output port 2, regardless of turns and defects. This path for the acoustic signal is a reconfigurable one-way channel, and transmission from 2 to 1 is zero. (b) Frequency dependence of the acoustic power transmission coefficient from 1 to 2 ($T_{1\rightarrow 2}$, blue curves) and from 2 to 1 ($T_{2\rightarrow 1}$, red curves) in the case of the lattice acoustic diode in panel (a) (solid lines), demonstrating broadband isolation. For comparison, a single trimer behaves like an imperfect acoustic diode operating only in the vicinity of a single frequency (dashed lines).

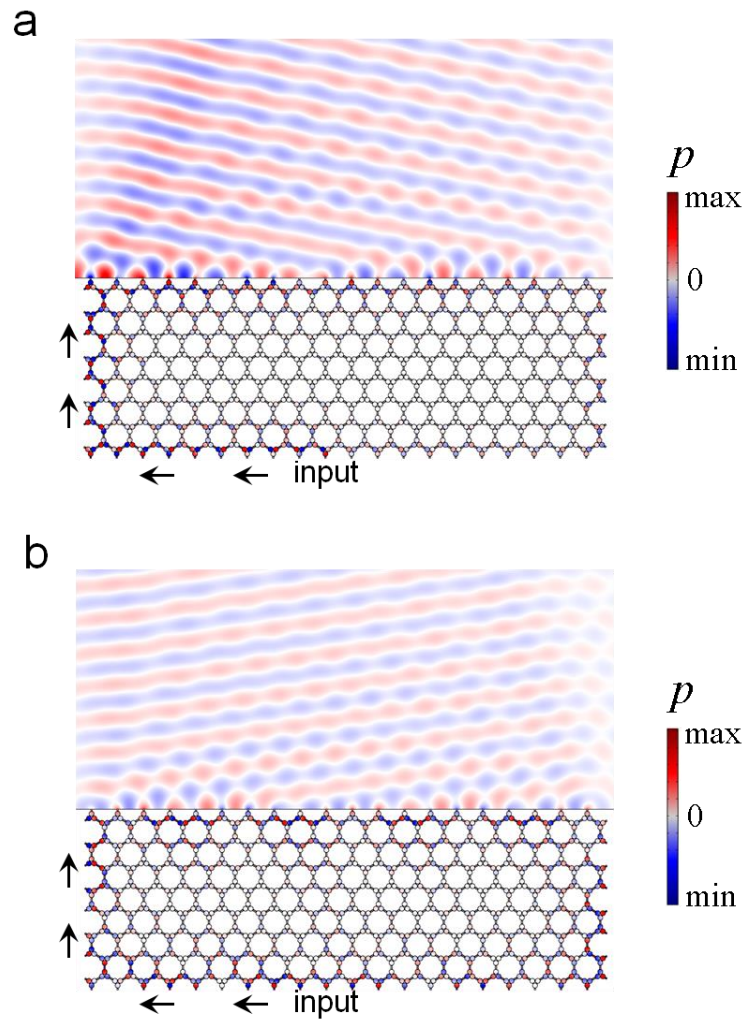

Figure 6: **Topologically protected leaky-wave acoustic antenna.** Topological edge states are robust also in the presence of radiative loss, enabling leaky-wave radiation from a crystal edge into a surrounding water domain, both in the forward (a) and backward (b) directions, for two different frequencies. Interestingly, such an acoustic leaky wave antenna possesses 100% efficiency, regardless of aperture truncation.

Topologically protected edge modes may also occur in the presence of radiative losses, when coupling to the radiation continuum takes place at the lattice boundary. This interesting possibility may be exploited in a variety of applications involving acoustic radiation, including underwater acoustic communications systems, acoustic detection and imaging. In Figure 6, we consider a topologically protected acoustic leaky-wave antenna built by allowing a crystal edge mode to radiate in water, while the bottom and side boundaries are hard-wall boundaries (e.g., interfaced with air). The input (feeding) point is located at the bottom edge. One of the practical challenges in conventional leaky-wave antennas is that, if the antenna is truncated before all the energy is radiated out, back-reflections cause the appearance of an unwanted beam in the specular direction, impacting directivity and sensitivity to noise. Conversely, as seen in the figure, the proposed topological leaky-wave antenna not only is capable of beam scanning the acoustic radiation both in the forward (Figure 6a) and backward directions (Figure 6b), by operating at different frequencies, but the radiation pattern remains highly directional even though the antenna is truncated before all the energy is radiated out, due to the intrinsic absence of a backward mode. Thanks to topological protection, all the power that is not radiated is re-routed at the back of the system and recycled. Different from conventional leaky-wave antennas, which lose their functionality and efficiency upon size reduction, radiation efficiency here is 100% regardless of truncation, and the antenna does not have impedance matching issues. It is also possible to emit at the opposite angle at the same frequency by simply reversing the lattice handedness (not shown for brevity). More importantly, this acoustic emitter fundamentally breaks the reciprocity constraint, dictated by time-reversal symmetry, based on which a good sound emitter is also a good receiver [53]. Due to the one-way properties of the edge mode responsible for radiation, an acoustic signal impinging on the open edge would not couple into the waveguide, realizing an acoustic emitter that is not prone to noise and interference from the background.

In summary, we have proven by first-principle simulations that time-modulated acoustic lattices represent an ideal platform to implement the anomalous physics of Floquet topological insulators, taking a pivotal step towards practical applications of topological acoustics. Using a simple but functional design with readily available acoustic materials, combined with an efficient low frequency on-site modulation scheme, we have demonstrated that topologically nontrivial properties for sound can be readily obtained in practice without requiring uniform phase of the modulation across the lattice, circumventing a practical issue that usually drastically complicates and obscures the applicability of time-modulated crystals. The topological state is only dictated by the local rotational pattern of the modulation phase that is delivered to every resonator, and is totally independent on its phase, which can be completely random from resonator to resonator. This work takes the idea of Floquet topological insulators from an abstract solid-state concept to a new fully-practical paradigm in acoustics with large application potential, paving the way to new acoustic systems leveraging the highly demanded features of topological insulators. The unprecedented broadband nonreciprocal response demonstrated here promises a new class of nonreciprocal elements – acoustic diodes and circulators – which possess unique characteristics of large bandwidth, insensitivity to impedance mismatch, and robustness to imperfections in the design implementation. While originally proposed in condensed matter and photonics, Floquet topological insulators may have a surprising twist, finding their first important applications in acoustics.

## Methods

**Tight binding model and stroboscopic evolution of acoustic Floquet topological insulators**

As described in the main text, a semi-analytical approach based on the effective Hamiltonian derived with the use of the stroboscopic evolution operator for the time-dependent tight-binding model is used. The Hamiltonian (1) in the main text has a time-modulated on-site part described by the matrix of Eq. (2), whose diagonal elements are eigenfrequencies of the cavities constituting the metamolecule. These diagonal elements vary with time to form a rotational pattern (i.e. $\omega_l(t) = \delta\omega \cos(\omega_l t + \phi_l)$, $\phi_l = \frac{2\pi}{3}l$, $l = 1,2,3$), and the off-diagonal components are kept constant. The off-site terms in Eq. (2) are described with a diagonal matrix, due to the numbering convention shown in Fig. 1SM.

We show in Figure 2SM the band structure for Hamiltonian (1) without time modulation. The resonant frequency $\omega_0$, the coupling parameter $\kappa$, and the hopping parameter $J$, are extracted from first-principles finite-element method (FEM) numerical simulations performed in COMSOL Multiphysics (Acoustic module). As expected, the band structure calculated from the tight-binding Hamiltonian (1) perfectly fits the band structure shown in the main text (Fig. 1C, blue bands), obtained from our numerical experiments.

Next, the time-independent stroboscopic effective Hamiltonian is found as $\widehat{\mathcal{H}}_{\text{eff}} = \frac{i}{T} \log \widehat{U}(T)$, where $\widehat{U}(t) = \mathcal{T} \exp\left[-i \int_0^t \widehat{\mathcal{H}}(t)\, dt\right]$ is the evolution operator evaluated over one period. The exponent of the integral in the evolution operator is calculated by numerical integration over one period $\widehat{U}(T) = \mathcal{T} \prod_n \exp\left(-i\Delta t \widehat{\mathcal{H}}(t_n)\right)$, where $t_n = n\Delta t \in [0, T)$. Here, $T$ is the period of modulation, and $\Delta t$ is the time step. Figure 3SM shows the band structure calculated numerically from the effective Hamiltonian $\widehat{\mathcal{H}}_{\text{eff}}$, both without modulation $\delta\omega = 0$ (blue bands) and with modulation $\delta\omega \neq 0$ (red bands). One can clearly see the opening of the complete acoustic band gap predicted by the tight binding approach, in agreement with the full-wave numerical results shown in the main text (Figs 1C and 1D).

To confirm the topological nature of the transition from the gapless to the gapped state, we calculated the Berry curvature $\Omega_n(\boldsymbol{k}) = \partial_{k_x} A_y - \partial_{k_y} A_x$, where $\mathbf{A} = -i\langle p_n | \partial_{\boldsymbol{k}} | p_n \rangle$ is the Berry connection, and $|p_n\rangle$ is the eigenstate of the effective Hamiltonian $\widehat{\mathcal{H}}_{\text{eff}}$. The Berry curvature for the four bands of interest calculated at every point in the Brillouin zone is plotted in Fig. 4SM. The subsequent integration of $\Omega_n(\mathbf{k})$ over the Brillion zone yielded the values of the Chern number $C_n = \{\pm 1, 0, 0, \mp 1\}$ for the clockwise/counterclockwise modulation of the phase in the trimer metamolecule. Note that the two other bands do not show any topological behavior and the Dirac degeneracy at K point insured by the lattice symmetry persists even in the presence of the modulation. Thus no topological band gap is observed for these bands and the corresponding states remain topologically trivial.

Finally, in order to demonstrate the presence of an edge state, a lattice supercell was modeled using the effective Hamiltonian obtained from the stroboscopic evolution operator. The supercell consists of 10 unit cells with a domain wall in the center of the cell, i.e. an abrupt flip of the modulation spin. After the time-dependent Hamiltonian for the supercell $\widehat{\mathcal{H}}^{\text{SC}}(t)$ was constructed, the procedure was identical to that of the single cell. The eigenmodes of the $\widehat{\mathcal{H}}_{\text{eff}}^{\text{SC}} = \frac{i}{T}\log \widehat{U}^{\text{SC}}(T)$, with $\widehat{U}^{\text{SC}}(t) = \mathcal{T}\exp\left[-i\int_0^t \widehat{\mathcal{H}}^{\text{SC}}(t)\,dt\right]$ being the supercell evolution operator, were numerically found. The resultant band structure of the supercell is shown in Fig. 5SM, and clearly reveals the presence of the four isolated bands (red lines) within the band gap region induced by the spatio-temporal phase modulation (bulk bands are shown in blue). The inspection of the eigenvectors confirms that these modes are localized to the domain wall, with the edge modes with the positive (negative) group velocity propagating along a RH/LH (LH/RH) boundary, thus confirming their one-way (chiral) character, which is in perfect agreement with the first-principle results presented in the main text.

**Finite-element frequency-domain simulations of acoustic Floquet topological insulators**

To perform a full-wave numerical experiment of our time-dependent lattice, we start by noticing that in order to modulate the cavity capacitance $C_0 = V_0 \beta_0$ by a relative amount $\delta C / C_0$, we can either modulate its volume by an amount $\delta V / V_0 = \delta C / C_0$, or equivalently the compressibility by an amount $\delta \beta / \beta_0 = \delta C / C_0$. In practice, it is naturally easier to modulate directly the volume, by compressing the cavities using piezoelectric actuators. Such a strategy indeed changes the volume without changing the compressibility, as long as linearity holds and Hookes law remains valid. However, for numerical purposes, the capacitance modulation is easier to model via a compressibility modulation, which has the same effect than the volume modulation, but avoids the complexity of a moving mesh and deformed geometry in a numerical code. Therefore, we start from first principles, i.e. Euler equation for the motion of a fluid particle and the conservation of mass:

$$\begin{cases} \nabla p(\mathbf{r},t) = -\dfrac{d}{dt}(\rho_0 \mathbf{u}(\mathbf{r},t)) \\ \nabla \cdot \mathbf{u}(\mathbf{r},t) = -\dfrac{d}{dt}(\beta(\mathbf{r},t) p(\mathbf{r},t)) \end{cases}, \qquad (3)$$

where $p(\mathbf{r},t)$ is the acoustic pressure, $\mathbf{u}(\mathbf{r},t)$ is the particle velocity, $\rho_0$ is the time independent density of the medium, and

$$\beta(\mathbf{r},t) = \beta_0(\mathbf{r}) + \delta\beta(\mathbf{r})\cos(\omega_m t - \varphi(\mathbf{r})) \qquad (4)$$

is the dynamically modulated compressibility of the structure. From Eq. (3), we get the wave equation $\Delta p(\mathbf{r},t) = \rho_0 \dfrac{d^2}{dt^2}(\beta(\mathbf{r},t) p(\mathbf{r},t))$, and use Floquet Bloch theorem in time domain, writing

$p(\mathbf{r},t) = \sum_n f_n(\mathbf{r})e^{i(\omega+n\omega_m)t}$. After a few calculation steps, we obtain the following infinite set of coupled time-independent differential equations for the harmonics $f_n(\mathbf{r})$:

$$\Delta f_n(\mathbf{r}) + \rho_0 \beta(\mathbf{r})(\omega+n\omega_m)^2 f_n(\mathbf{r}) = -\frac{1}{2}\rho_0 \delta\beta_0(\mathbf{r})(\omega+n\omega_m)^2 \left(f_{n-1}(\mathbf{r})e^{-i\varphi(\mathbf{r})} + f_{n+1}(\mathbf{r})e^{i\varphi(\mathbf{r})}\right). \quad (5)$$

In our numerical simulations, the relative weakness of the modulation allows solutions to converge quicky, and the system (5) can be truncated to the five harmonics $n = \{-2,-1,0,1,2\}$ that dominate the field in the frequency range of interest. The corresponding equations (5) have been transformed into weak form and directly introduced into the finite-element solver COMSOL Multiphysics (Acoustic Module) where fully coupled equation system was directly solved in the (quasi-) frequency domain.

**Acknowledgments**

This work was partially supported by the National Science Foundation and the Department of Defense.

**Author information**

**Contributions**

All authors contributed extensively to the analytical derivations, numerical calculations, and the preparation of the manuscript.

**Additional information**

Correspondence and requests for materials should be addressed to Andrea Alù alu@mail.utexas.edu, Alexander Khanikaev akhanikaev@qc.cuny.edu.

**Competing financial interests**

The authors declare no competing financial interests.

# Supplementary information

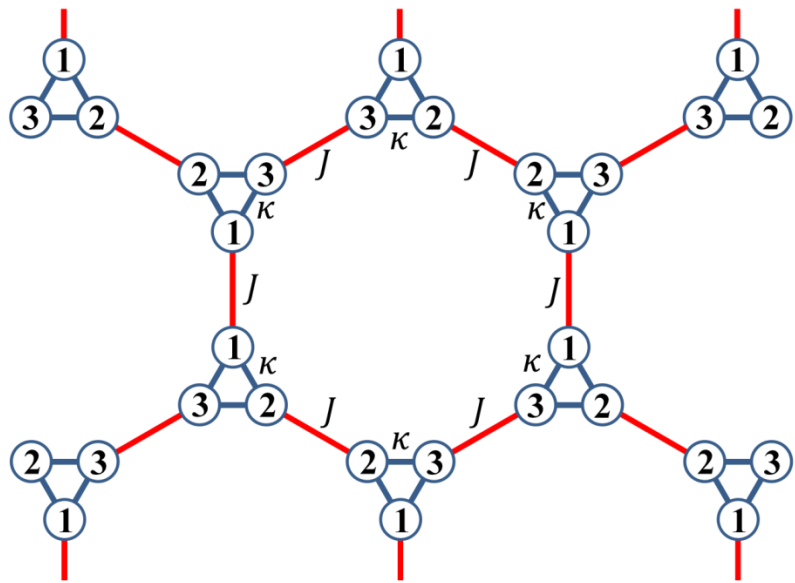

**Figure 1SM| Hexagonal lattice of the trimer metamolecules.** The metamolecules are composed of acoustic cavities indexed as 1, 2, and 3. Blue links show inner connections between the cavities with the strength of coupling (on-site hopping) $\kappa$ and the red lines show the nearest neighbor (off-site) hopping $J$ between the metamolecules.

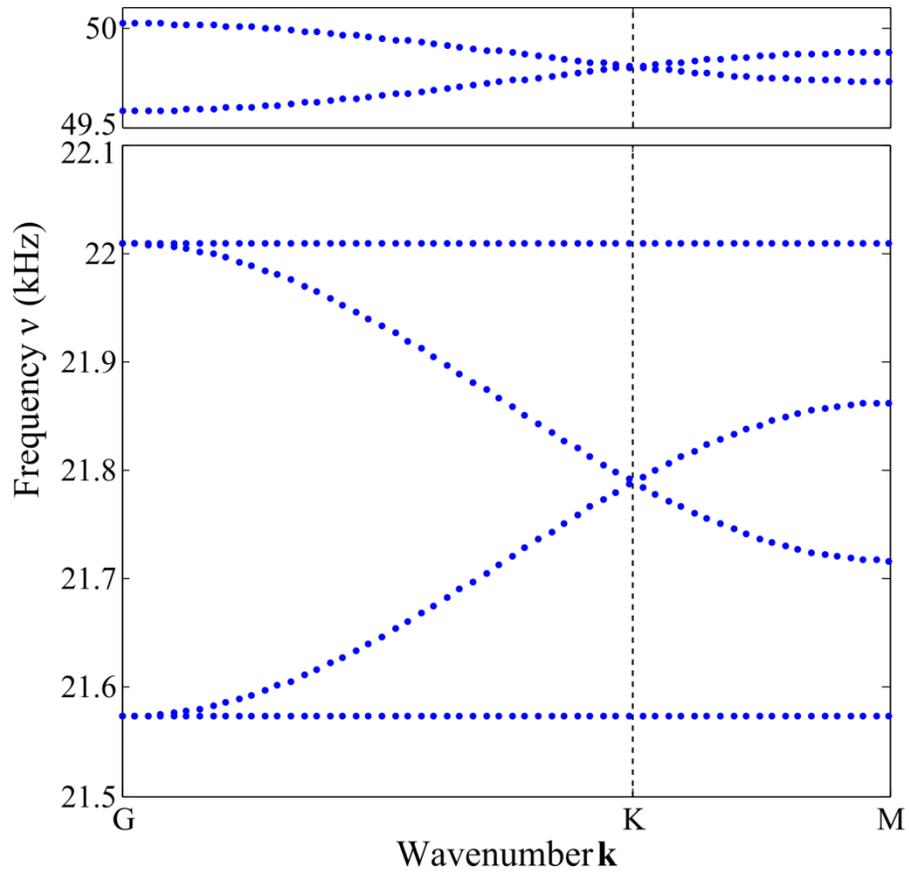

**Figure 2SM| Acoustic band structure of the tight binding Hamiltonian Eq.(1) without temporal modulation** ($\delta\omega = 0$). The parameters used are $\nu_0 = \omega_0/2\pi =31.13$ kHz, $\kappa = 0.3\nu_0$, and $J = 0.7 \times 10^{-2}\nu_0$.

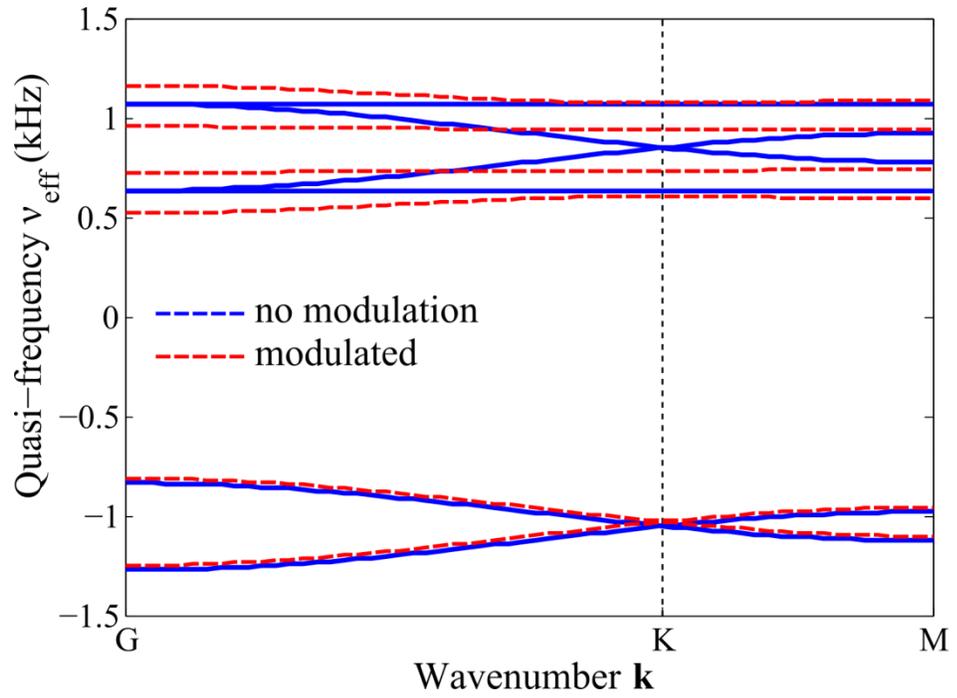

**Figure 3SM| Acoustic band structure of obtained from the effective Hamiltonian $\widehat{\mathcal{H}}_{\text{eff}}$ describing the stroboscopic evolution of the system.** Cases without the temporal modulation ($\delta\nu = 0$, blue solid lines) and with the temporal modulation ($\delta\nu = 0.036\nu_0$, red dashed lines) are shown. The parameters are the same as in Fig. 2SM, and the modulation frequency $\nu_m = \omega_m/2\pi = 3$ kHz.

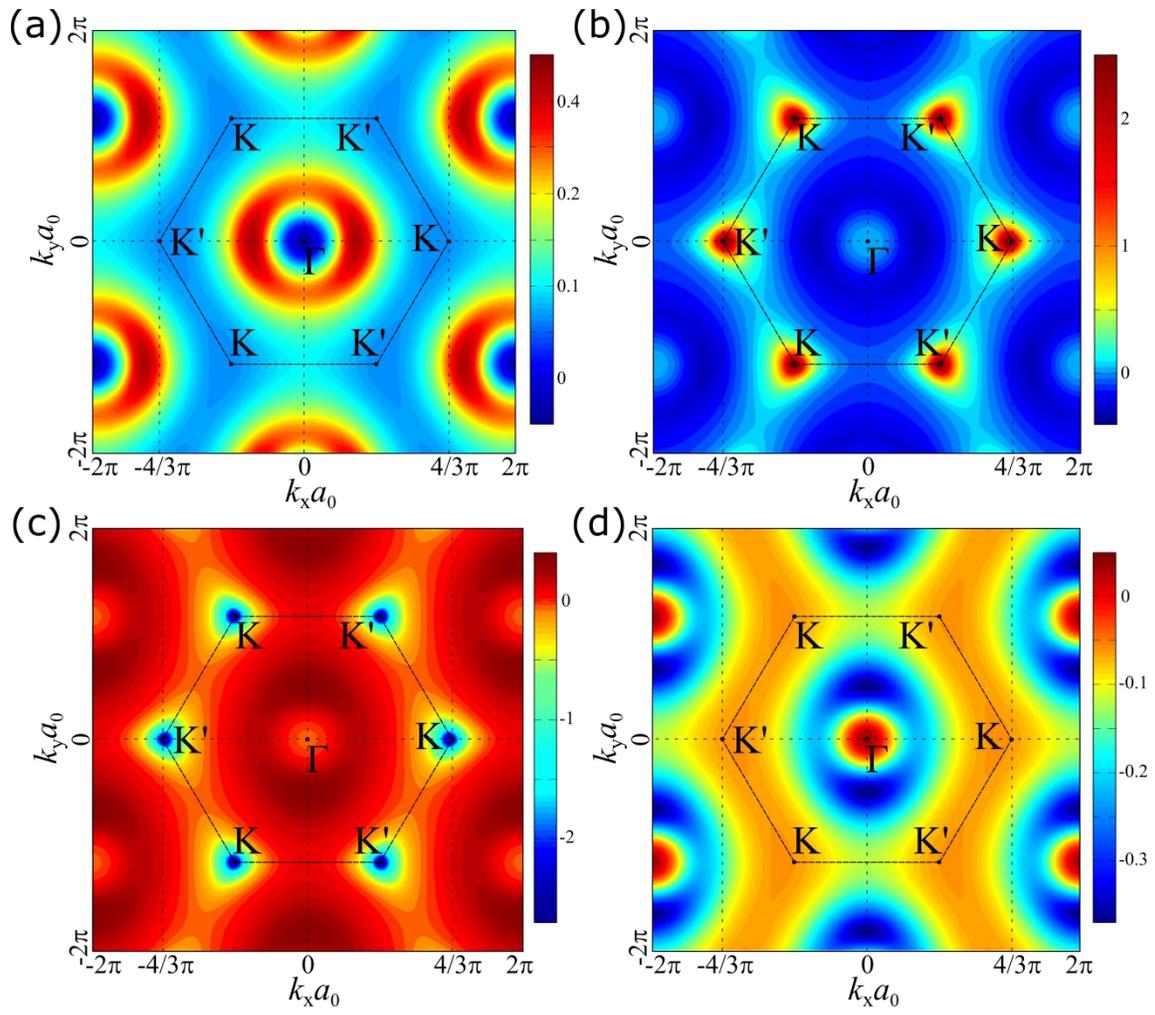

**Figure 4SM| Berry curvature for the four topological bands of acoustic Floquet topological insulator.**
The Berry curvature is calculated from the effective Hamiltonian $\hat{\mathcal{H}}_{\text{eff}}$ for the lattice parameters given in Figs. (2) and (3), in the presence of modulation. Subplots (a)-(d) correspond to the four bands, from lower to higher frequency, respectively.

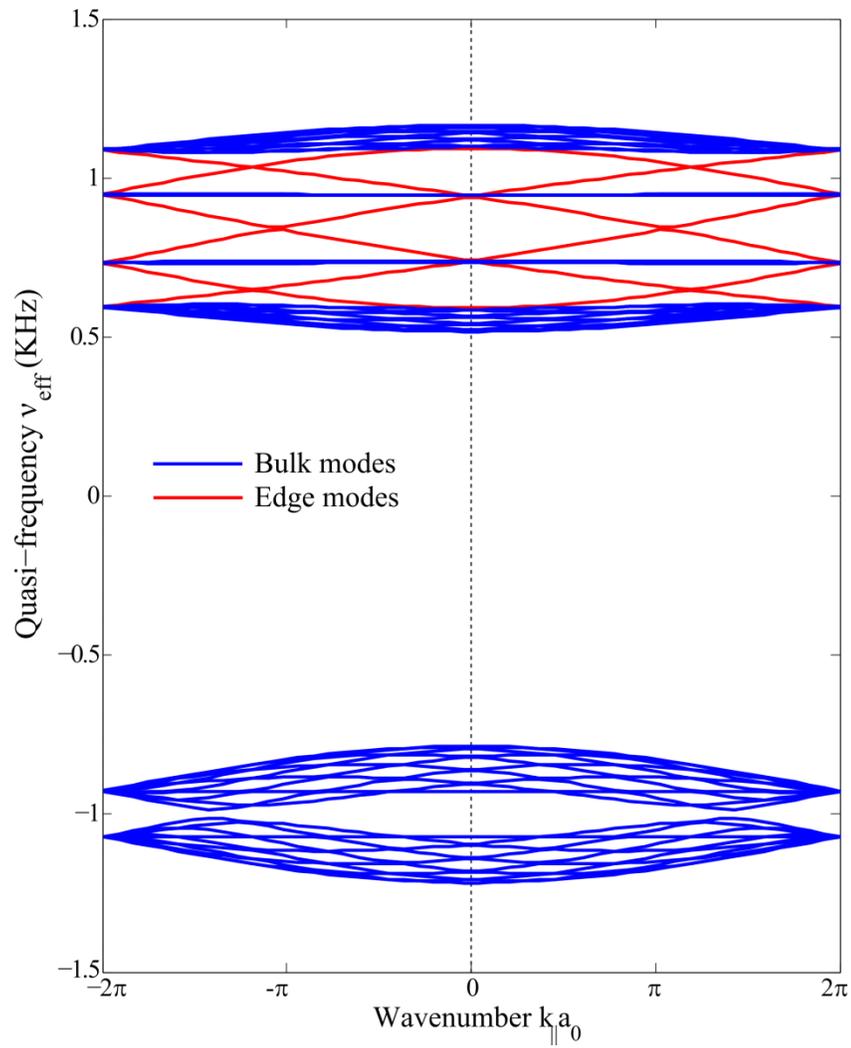

**Figure 5SM| Acoustic band structure of the 10x1 supercell with domain wall at the center, obtained from the effective Hamiltonian of the stroboscopic evolution.** Blue and red lines correspond to the bulk and edge modes, respectively. The parameters used are the same as above.